\documentclass{emulateapj}
\slugcomment{}

\shorttitle{Optimal filter systems for photo-z}
\shortauthors{Ben\'\i tez et al.}

\begin{document}

\title{Optimal filter systems for photometric redshift estimation}

\author{N. Ben\'\i tez\altaffilmark{1,2},  
	M. Moles\altaffilmark{1},  
   	J.A.L. Aguerri\altaffilmark{3}, 
        E. Alfaro\altaffilmark{1}, 
        T. Broadhurst\altaffilmark{4}, 
        J. Cabrera-Ca\~no\altaffilmark{5}, 
        F.J. Castander\altaffilmark{6}, 
        J. Cepa\altaffilmark{3}, 
        M. Cervi\~no\altaffilmark{1}, 
	D. Crist\'obal-Hornillos\altaffilmark{1}, 
        A. Fern\'andez-Soto\altaffilmark{7,8}, 
        R.M. Gonz\'alez Delgado\altaffilmark{1}, 
        L. Infante\altaffilmark{9}, 
        I. M\'arquez\altaffilmark{1}, 
        V.J. Mart\'\i nez\altaffilmark{7}, 
        J. Masegosa\altaffilmark{1}, 
        A. Del Olmo\altaffilmark{1}, 
        J. Perea\altaffilmark{1}, 
        F. Prada\altaffilmark{1}, 
        J.M. Quintana\altaffilmark{1} and  
	S.F. S\'anchez\altaffilmark{10}}

\altaffiltext{1}{
Instituto de Astrof\'\i sica de Andaluc\'\i a (CSIC), Apdo. 3044, 18008 Granada, Spain}

\altaffiltext{2}{IFF(CSIC), C/Serrano 113-bis, 28005 Madrid, Spain}

\altaffiltext{3}{
Instituto de Astrof\'\i sica de Canarias, 
V\'\i a L\'actea s/n, La Laguna, Tenerife 38200, Spain}

\altaffiltext{4}{
The Department of Astronomy and Astrophysics, Tel-Aviv University, Israel}

\altaffiltext{5}{
Departamento de F\'\i sica At\'omica, 
Molecular y Nuclear, Facultad de F\'\i sica, Universidad de Sevilla, Spain} 

\altaffiltext{6}{
Institut d'Estudis Espacials de Catalunya/CSIC, 
Gran Capit\'a 2-4, 
08034 Barcelona, Spain.} 

\altaffiltext{7}{
Obs. Ast. Univ.Valencia, Edificio de Institutos, 
Pol\'\i gono de la Coma s/n, Paterna-46980-Valencia,Spain
}

\altaffiltext{8}{
Instituto de F\'\i sica de Cantabria (CSIC), 39005 Santander, Spain}

\altaffiltext{9}{
Departamento de Astronom\'\i a y 
Astrof\'\i sica, Pontificia Universidad Cat\'olica de Chile.
Campus San Joaqu\'\i n. Vicu\~na Mackenna 4860 Casilla 306. Santiago 22. Chile}

\altaffiltext{10}{Centro Astron\'omico Hispano-Alem\'an, Almer\'\i a, Spain}

\begin{abstract}

   In the next years, several cosmological surveys will rely on imaging data 
to estimate the redshift of galaxies, using traditional filter systems with 
$4-5$ optical broad bands; narrower filters improve the spectral resolution, but strongly reduce the total system throughput. We explore how photometric redshift performance depends on the number of filters $n_f$, characterizing the survey depth by the fraction of galaxies with unambiguous redshift estimates. For a combination of total exposure time and 
telescope imaging area  of $270$ hrs m$^2$, $4-5$ filter systems perform  significantly worse, both in completeness depth and precision, than systems 
with $n_f\gtrsim 8$ filters. Our results suggest that for low $n_f$, 
the color-redshift degeneracies overwhelm the improvements 
in photometric depth, and that even at higher $n_f$, 
the effective {\it photometric redshift depth} decreases much 
more slowly with filter width than naively expected from 
the reduction in $S/N$. Adding near-IR observations improves the 
performance of low $n_f$ systems, but still the system 
which maximizes the photometric redshift completeness is formed 
by $9$ filters with logarithmically increasing bandwidth 
(constant resolution) and half-band overlap, reaching $\sim 0.7$~mag deeper, 
with $10\%$ better redshift precision, than $4-5$ filter systems. 
A system with $20$ constant-width, non-overlapping filters reaches only $\sim 0.1$~mag shallower than $4-5$ filter systems, but has a precision almost 3 times better,  $\delta z = 0.014(1+z)$ vs $\delta z = 0.042(1+z)$. 
We briefly discuss a practical implementation of such a photometric 
system: the ALHAMBRA survey.

\end{abstract}

\maketitle

\section{Introduction}

  Photometric redshift estimation is not a new technique (Baum 1962,
Loh \& Spillar 1986, Koo 1985, Connolly et al. 1995; see Koo 1999
for a history of the method), but it has considerably developed in
the last decade, especially following the Hubble Deep Field (HDF) observations
(Williams 1996, Casertano et al. 2000), which provided catalogs with 
excellent photometric quality and abundant spectroscopic redshift coverage. 
This allowed astronomers to thoroughly test standard photo-z techniques 
and try new approaches
(Gwyn \& Hartwick 1996, Lanzetta, Yahil \& Fern\'{a}ndez-Soto 1997,
Sawicki, Lin \& Yee 1997, Fern\'{a}ndez-Soto, Lanzetta \& Yahil 1999,
Brunner et al. 1997, Ben\'{\i}tez et al. 1999, Ben\'{\i}tez 2000, 
Bolzonella, Miralles \& Pell\'{o} 2000).

  As Hickson, Gibson \& Callaghan (1994) first showed, multiband narrow filters can be 
much more efficient to obtain redshifts than spectroscopy if the large area of the imaging
cameras is factored in. Several photometric surveys, using different filter systems,
have been proposed or implemented in the last decade: the
UBC-NASA survey (Hickson \& Mulrooney 1998), CADIS (Wolf et al.
2001b), COMBO-17 (Wolf et al. 2001a), COSMOS-21 (Taniguchi 2004), ALHAMBRA (Moles et al. 2008),
DES (DES collaboration 2005), LSST (Tyson 2006), PanStarrs (Kaiser 2007), 
VST (Arnaboldi et al. 2007),  and PAU (Ben\'\i tez et al. 2008). These surveys 
represent powerful  alternatives 
to deep spectroscopic surveys like DEEP2 (Davis et al. 2003), 
VVDS (Le F\`{e}vre et al. 2003), or BOSS (Schlegel et al. 2007) 
at least for those scientific goals which only require limited redshift 
accuracy and low resolution spectral information.

  However at least three of the imaging surveys 
(DES, LSST, PanStarrs) will work with photometric 
systems with $4-5$ optical broadband filters, similar to those traditionally used 
in Astronomy. It is obvious that using more, narrower filters, for a fixed 
exposure time, will significantly 
sacrifice photometric depth. 
However, photometric depth is not equivalent to 
{\it photometric redshift} 
depth. The fewer the filters, the more prone the system is to color-redshift degeneracies; 
these make impossible to unambiguously determine the redshift for a galaxy, even 
if observed at relatively high $S/N$. The Hubble Ultra Deep Field (Beckwith et al. 2003) 
offers a good example. Despite the fact that the limiting magnitude in the HUDF is 
$0.9-1.4$ mag 
deeper than the HDF, the lack of a 
$U-$band filter in the HUDF makes the photometric redshift depth of 
both fields similar (Coe et al. 2006). 

  This letters explores 
the impact on photometric redshift performance of factors as the number of filters $n_{f}$, constant vs. 
logarithmically increasing 
bandwidth, half-band overlaps, and near-IR observations. 
We also briefly discuss a practical implementation of a medium band filter 
system: the ALHAMBRA survey. 

\section{Simulations}

\subsection{Description of the filter systems}

 We assume that filters are almost ``top-hat'', 
with a transmission which is constant in the central part and steeply falls on the 
edges, formed by half-Gaussian wings with a HWHM of $\sim 15$ \AA.
Although somewhat idealized, this is very similar to the characteristics of the 
filters provided by BARR Associates for the ALHAMBRA survey. We consider four type of 
photometric systems, depending on whether they have constant or logarithmically increasing 
($\Delta\lambda\sim\lambda$) bandwidth, and whether they have half-width $\Delta\lambda/2$ 
overlaps or just a minimal overlap corresponding to the filter wings. 
The filters cover the $3400-9800$\AA interval. 
Fig. 1. shows examples of the 4 types of filter systems considered.

\begin{figure}
\caption{Example of the four types of filter sets considered, 
each with 11 filters. We represent the filter transmissions without 
taking into account the CCD or the telescope+optics transmission, 
factors which are later included in the photometric noise 
estimation to produce realistic photometric measurements.
We slightly increase the height of the filters with wavelength, and 
alternate colors in successive filters to help visualization.}
\plotone{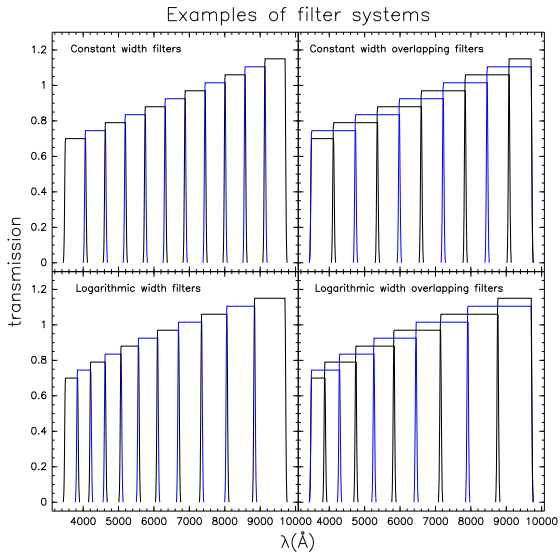}
\end{figure}

\subsection{Mock catalogs}

 To generate realistic galaxy distributions, we use the HDF catalogs
of Fern\'{a}ndez-Soto, Lanzetta \& Yahil (1999) and Yahata et al.
(2000). It is possible to obtain accurate and reliable photometric
redshifts and type classifications, based on the extended Coleman,
Wu and Weedman (1980) set for these galaxies up to $I_{AB}\approx27$
(Ben\'{\i}tez 2000, Fern\'{a}ndez-Soto et al. 2001). We use the
distribution of $I_{AB}$, Bayesian spectral type $t_{b}$ and Bayesian
photometric redshift $z_{b}$ obtained from these catalogs with the
software BPZ (http://acs.pha.jhu.edu/$\sim$txitxo, Ben\'\i tez 2000) and 
the empirically calibrated spectral library of Ben\'\i tez et al. 2004.   
The original input catalogs contain 822 galaxies with $I_{AB}<26$. 
For each filter set combination, our simulation includes $5,000$ galaxies, 
generated by randomly choosing objects from the HDF catalogs. The depth 
reached by our simulations (up to $I\sim 25$) precludes using an input catalog 
based on spectroscopic redshifts, which would not be complete enough at 
those magnitudes. Since the accuracy 
of the input photometric redshifts is $\approx 0.06(1+z)$, we perturb them 
by a similar, randomly distributed amount to produce a more realistic 
redshift distribution.

 We assume a constant total exposure time of $100,000$s (28hrs) per pointing,
and vary the number of filters $n_f$. The average exposure is thus 
$100,000/n_{f}$. Instead of using a fixed observing time per filter we 
distribute the exposures trying to reach constant $S/N$ 
for a same AB magnitude, with two constraints: the minimal exposure
time per filter is, for practical reasons, at least $2,500$s, and we
do not expose more than twice the length of the average exposure, 
i.e. $2\times100,000/n_{f}$ to avoid spending too much of the total time 
on the filters which are less efficient. For wavelengths $\lambda>8000$\AA 
the efficiency is so low that we set this limit to be $100,000/n_{f}$.

  Using this information, we generate the expected magnitudes in all
the filter systems described in the above section using functions 
included in the BPZ package. For $\sim 300$\AA~ filters the corresponding $5\sigma$ 
limiting magnitudes are $m_{AB}\sim 25$ bluewards of $8000$\AA~ and quickly 
degrades to $m_{AB} \sim 23$ at $9500$\AA. As it is obvious, for a fixed total 
exposure time, the limiting flux in a filter will be roughly inversely proportional 
to the square root of the filter width $\Delta\lambda$. 

  As a last step we add random photometric noise 
whose amplitude has been estimated using the WHT exposure time calculator\footnote{
Based on the SIGNAL software, by Chris Benn} scaling it to a 3.5m telescope. The 
product of exposure time by telescope area is $\sim 270$ hrs m$^2$, and therefore 
the results obtained here are equivalent to e.g. what could be obtained in a few hours of 
total exposure time with a $8-10$m telescope. 

 We also scale the $S/N$ as a function of the magnitude of the galaxies. We do not take 
into account the dependence of size, etc. with magnitude. These are second order effects
which will have a similar impact on all the filter systems and therefore are
not expected to significantly affect the comparison among them.

\section{Comparison among different filter systems}

  The photometric redshifts for the mock catalogs are estimated
using the BPZ package. The software provides a Bayesian estimate of
the redshift and a spectral type classification. The expected 
reliability of the photo-z can be gauged through the Bayesian odds. 
The value of this parameter corresponds to the amount of redshift 
probability concentrated on a $\pm 0.2(1+z)$ region around the 
probability maximum. Low values of the odds indicate a multimodal 
or very extended, little informative $p(z)$, indicating that 
the photometric information is insufficient to obtain a unambiguous estimate 
of the galaxy redshift. By selecting objects with high odds, e.g. 
$\geq 0.99$, one can produce highly reliable samples (Ben\'\i tez 2000), 
with very good redshift accuracy and a very low rate ($\lesssim 2\%$) of 
``catastrophic'' outliers. Therefore, it is possible to accurately 
characterize the effective completeness of a photometric redshift 
catalog by using the amount of galaxies with odds above 
a certain threshold, which tells us how many galaxies 
we can expect to have meaningful, univocal photometric redshifts.

 For a set-up with a total exposure time $T$ and total number of filters $n_f$, 
the signal--to--noise in an individual filter, assuming that we are limited by the 
sky background, would roughly change as $S/N_i \propto 1/{n_f}$. A way of comparing  
depths across different systems is the $S/N_B$ in a fixed width band 
(obtained by combining all the individual filters included in that band):  
$S/N_B \propto 1/\sqrt{n_f}$. Therefore, by increasing the number of filters we would 
expect the effective limiting magnitude at a fixed $S/N$ level 
to diminish quite drastically, as $m_{lim} =$ const$ + 2.5 \log(\sqrt{n_f})$, 
e.g. equivalent to the loss of a full magnitude going from $4$ to $25$ filters.  

  Fig. 2 describes how the $80\%$ completeness magnitude limit 
behaves for each of the filter systems. We see that for contiguous 
filters, the completeness depth sinks fast for $n_f < 8$, and 
that the optimum number of filters is $n_f\sim 12$, 
after which the effective completeness magnitude decreases, but 
much more slowly than expected from the change in the photometric limiting magnitude. 
This shows that for systems with low $n_f$, the color-redshift degeneracies 
introduced by an insufficient wavelength resolution dominate over the improvement 
in $S/N$ achieved by the increased filter width. 

 Fig. 3 shows what happens when we add moderately deep near-IR 
observations with $5\sigma$ limiting (Vega) magnitudes of $J=22.4,H=21.2,K=20.4$. 
There is a very significant, almost $\sim 0.4$ mag increase in the completeness 
magnitude, and the behavior of the low $n_f$ systems relatively improves, 
but still the most efficient overall performer 
is a logarithmically increasing bandwidth, half-band 
overlapping system with $9$ filters, which reaches a 
completeness limit $\sim 0.7$ mag deeper than a typical $4-5$ 
filter system with the same exposure time, while having a $10\%$ better accuracy. 

\begin{figure}
\caption{
 Effective $80\%$ completeness magnitude, corresponding to 
the magnitude at which the accumulated number of objects 
$N(<m_{0.99})$ with 
 Bayesian $odds\geq 0.99$ is $80\%$ of the total number of objects 
$N(<m)$, a good measure of the effective depth of 
a survey. The blue dotted line illustrates how the completeness 
magnitude would change with filter number if it mimicked the behavior 
of the photometric limiting magnitude.}   
\plotone{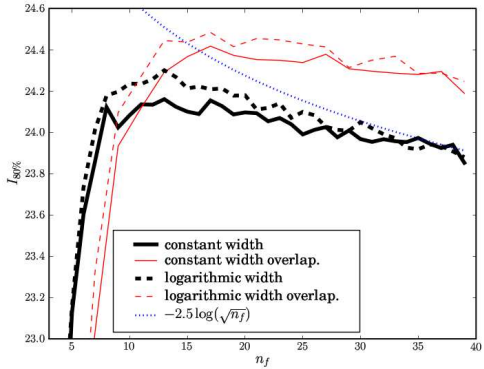}

\end{figure}

\begin{figure}
\caption{
Same as previous figure, but including moderately deep near-IR 
observations (see text for details)}
\plotone{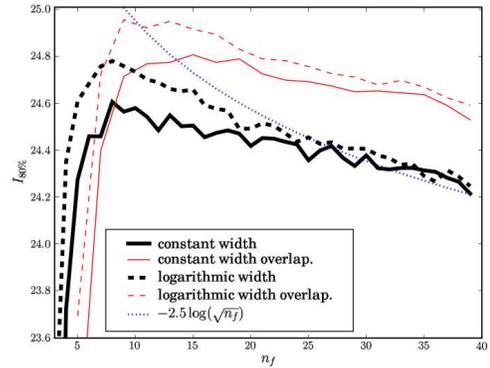}
\end{figure}

  Another obvious quantity to consider is the accuracy of the photometric 
redshifts for the high $odds$ sample, estimated using the 
$rms$ of the quantity $\Delta z/(1+z)=(z-z_{b})/(1+z)$, plotted 
in Fig. 4. Here we see that, as expected, the redshift precision 
quickly and monotonously improves with $n_f$ and that adjacent filter 
systems perform much better than overlapping ones. From Fig. 3 we can see 
that a adjacent system with $n_f=20$ reaches a completeness 
depth similar to traditional systems with $n_f=5$, but an accuracy significantly 
better: $0.015(1+z)$ vs $0.04(1+z)$.

\begin{figure}
\caption{Dependence of the rms of quantity $(z-z_{b})/(1+z)$ for those galaxies
with $Odds>0.99$ as a function of the number of filters for the four
types of filter system considered in the paper and 
including near-IR observations (see text for details).
}  
\plotone{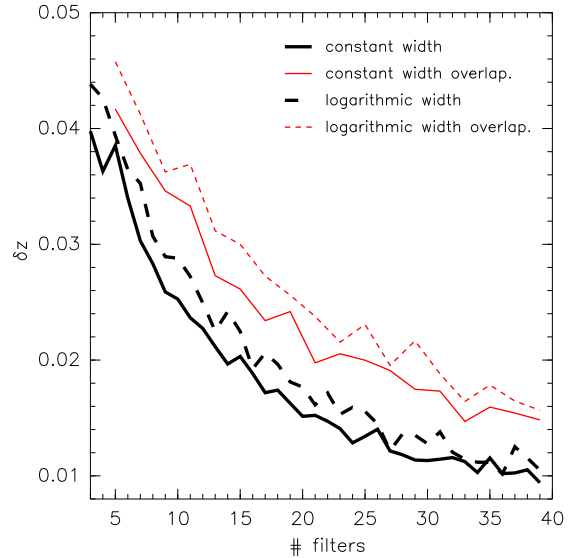}
\end{figure}

\section{The ALHAMBRA Survey}

  Considering the above results, plus additional requirements on emission line detection,  
the ALHAMBRA (Advanced, Large, Homogeneous Area, 
Medium Band Redshift Astronomical) survey decided to use a constant-width, non overlapping 
filter system, complemented with near-IR observations. ALHAMBRA had 
an additional requirement, the detection of a large fraction of galaxies with 
emission lines, which favored the choice of $310\AA$ filters. The ALHAMBRA 
$3\sigma$ rest-frame detection limits for a typical AB$\approx 23$ galaxy are  
EW(H$\alpha$)$>28$\AA~ 
out to $z\approx0.45$, and EW(O{\rm{II})$>16$\AA~ out to 
$z\approx1.55$. 
From comparison with Hippelein et al. 2003, 
ALHAMBRA expects to detect $\approx50$\% of the H$\alpha$ 
emitters at $z\approx0.25$, and $\approx80$\% of the O{\rm{II}objects to 
$z\approx1.2$; since $80\%$ of field galaxies at those redshifts have 
detectable emission lines (Tresse \& Maddox 1998) we expect to detect lines 
for a large fraction of our whole sample.   


  The survey is imaging 4 sq. degrees with the camera LAICA at the Calar Alto 3.5m telescope and also 
obtaining deep $JHK$ observations with Omega2000 at the same telescope.The survey characteristics, 
scientific goals and preliminary results are described in detail in Moles et al. (2008). A good test 
of the simulations presented in this paper is a comparison with ALHAMBRA. The mock catalogs 
predict that, with 20 filters, ALHAMBRA should be able to reach a precision of 
$\delta z/(1+z)\approx 0.014$ for $I\lesssim 24$ galaxies. Preliminary results show 
that the measured redshift error (Moles et al. 2008) is similar or less than $0.015$, 
supporting the validity of the simulations presented in this paper. 

\section{Conclusions}


  We explore the performance of four different uniform 
filter systems with constant and logarithmically increasing 
($\Delta\lambda\propto\lambda$) widths, 
and with half-width $\Delta\lambda/2$ overlaps or just a minimal overlap corresponding to the 
filter wings, and use, as a measure of survey effective depth, the fraction of galaxies with 
a compact, unimodal probability redshift distributions as a function of magnitude. 
Our simulations employ a realistic input catalog, based on HDF photometric redshifts, 
and correspond to a combination of total exposure time and 
telescope area of $270$ hrs m$^2$. We find that traditional $4-5$ optical filter systems clearly underperform, 
both in terms of completeness magnitude limit and precision, systems with $n_f\gtrsim 8$ filters. 

  Our results suggest that for low $n_f$, the effect of color-redshift degeneracies 
dominates the advantages of increased photometric depth, 
and that even at higher $n_f$, the effective {\it photometric redshift depth} decreases much 
more slowly with filter width than naively expected from 
the reduction in $S/N$. Adding near-IR observations increases the overall depth, 
alleviating color-redshift degeneracies and improving the relative performance 
of low $n_f$ systems. However the optimum performance still comes from a system with $9$ filters 
with logarithmically increasing bandwidth (constant resolution) and 
half-band overlap, which reaches $\sim 0.7$mag deeper, with $10\%$ better redshift 
precision, than $4-5$ filter systems. For many scientific applications, 
which require both precision and depth, the use of 
$>15$ medium band filters is clearly advantageous. A system with $20$ constant-width, 
non-overlapping filters reaches only $\sim 0.1$mag shallower than $4-5$ filter systems, 
but has a precision almost 3 times better,  $\delta z = 0.014(1+z)$ vs $\delta z = 0.042(1+z)$, 
as a practical implementation of such a system, the ALHAMBRA survey, shows. 

 Since it is well known that color-redshift degeneracies worsen 
with magnitude depth, it can be expected that the relative 
decoupling between photometric depth and photometric redshift depth described here 
will be more significant for surveys which reach fainter limits than 
those considered in our simulations, and less important for shallower observations, 
where the color/redshift degeneracies are less of a problem. In any case, 
future projects will have to seek an optimum number of filters 
based on their particular observing parameters and science goals.

\acknowledgements
  This work has been supported by the CONSOLIDER AYA2006-14056, 
and the proyecto intramural del CSIC 200750I003.

\end{document}